\documentclass[reqno,12pt]{amsart}

\usepackage{enumerate}
\usepackage[margin=1in]{geometry}
\usepackage{enumitem}
\usepackage{ifpdf}
\usepackage{amsmath}
\usepackage{amsfonts}
\usepackage{amssymb}
\usepackage[foot]{amsaddr}
\usepackage{amsthm}
\usepackage{mathrsfs}
\usepackage{amsrefs}
\usepackage[titletoc,title]{appendix}
\usepackage[dotinlabels]{titletoc}
\usepackage[all,cmtip]{xy}
\usepackage{nicefrac}
\usepackage{float}
\usepackage{verbatim}
\usepackage[multiple]{footmisc}
\usepackage{mathdots}
\usepackage{dcolumn}
\usepackage[hang,small,bf]{caption}    % fancy captions
\usepackage[overload]{textcase} 
\usepackage{graphicx}\usepackage{subcaption}
\usepackage[hyperfootnotes=false]{hyperref}

\newlength\Li \newlength\Lii 
\newcommand{\ignore}[1]{}

\theoremstyle{definition}

\theoremstyle{remark}

\makeatletter
\newcommand*{\rom}[1]{\expandafter\@slowromancap\romannumeral #1@}
\makeatother

\newtheoremstyle{defnopunct}
{3pt}
{3pt}
{}
{}
{\bfseries}
{ }
{.5em}
{}

\usepackage{etoolbox}
\let\bbordermatrix\bordermatrix
\patchcmd{\bbordermatrix}{8.75}{4.75}{}{}
\patchcmd{\bbordermatrix}{\left(}{\left[}{}{}
\patchcmd{\bbordermatrix}{\right)}{\right]}{}{}

\makeatletter
\patchcmd{\subsection}{\bfseries}{\itshape}{}{}
\patchcmd{\@sect}{\@addpunct.}{}{}{}
\makeatother

\author{David A. Meyer$^\dag$, Asif Shakeel$^\dag$}
\address{$^\dag$Department of Mathematics, University of California, San Diego, La Jolla, CA 92093-0112, USA}
\email{dmeyer@math.ucsd.edu, ashakeel@ucsd.edu}
\title [Mobility Trajectories from Network-Driven Markov Dynamics]{Mobility Trajectories from Network-Driven Markov Dynamics}

\begin{document}

\begin{abstract}
We present a generative model of human mobility in which trajectories arise as realizations of a prescribed, time-dependent Markov dynamics defined on a spatial interaction network. The model constructs a hierarchical routing structure with hubs, corridors, feeder paths, and metro links, and specifies transition matrices using gravity-type distance decay combined with externally imposed temporal schedules and directional biases. Population mass evolves as indistinguishable, memoryless movers performing a single transition per time step.

When aggregated, the resulting trajectories reproduce structured origin-destination flows that reflect network geometry, temporal modulation, and connectivity constraints. By applying the Perron-Frobenius theorem to the daily evolution operator, we identify a unique periodic invariant population distribution that serves as a natural non-transient reference state. We verify consistency between trajectory-level realizations and multi-step Markov dynamics, showing that discrepancies are entirely attributable to finite-population sampling. The framework provides a network-centric, privacy-preserving approach to generating mobility trajectories and studying time-elapsed flow structure without invoking individual-level behavioral assumptions.
\end{abstract}


\maketitle

 \section{Introduction} \label{sec:intro}
Human mobility modeling has traditionally relied on individual-level trajectories obtained from GPS traces, mobile phone records, or travel diaries. Early empirical studies demonstrated strong regularities in individual movement, including heavy-tailed displacement distributions and recurrent visitation patterns, suggesting a high degree of predictability in human mobility \cite{bhg:slht,ghb:uihmp}. These findings motivated trajectory-based models that explicitly encode individual behavior over time.

A central class of such models is the exploration–preferential return framework introduced by Song et al. \cite{song2010modelling,song2010}, in which agents alternate between exploring new locations and returning to previously visited ones with reinforcement proportional to visitation frequency. Variants of this approach incorporate circadian rhythms, activity schedules, or purpose-driven travel, and are designed to reproduce empirical statistics such as visitation heterogeneity and radius of gyration \cite{bbgj:hmma}. These models are inherently non-Markovian: transition probabilities depend on the full visitation history of each agent, and agents are therefore not interchangeable.

In parallel, flow-based mobility models focus on aggregate origin–destination patterns rather than individual trajectories. Gravity models and the radiation model describe population-level flows as functions of distance, population, or intervening opportunities \cite{simini2012}. While successful at reproducing large-scale mobility patterns, these approaches are essentially static: they specify expected flows over a given interval but do not define an explicit dynamical process evolving in time, nor do they generate realizations that can be interpreted as paths.

Viewed dynamically, aggregate origin–destination flows naturally define a stochastic process on a network. When interpreted as time-dependent Markov transition matrices, such flows specify how population mass redistributes over successive time steps. In most existing applications, however, these Markov dynamics are either assumed time-homogeneous, left implicit, or inferred retrospectively from individual trajectory data \cite{bbgj:hmma}, making the stochastic process primarily descriptive rather than prescribed.

In this paper, we reverse this perspective and treat the Markov dynamics itself as the primary modeling object. We prescribe a time-dependent, network-driven Markov process directly from spatial geometry, hierarchical connectivity, and externally imposed temporal biases, rather than inferring transition probabilities from individual trajectory data. Synthetic mobility trajectories are then realized as memoryless, interchangeable movers evolving under these prescribed dynamics. By construction, the resulting trajectories are consistent with aggregate origin–destination (OD) flow constraints and do not encode individual identities, preferences, or activity schedules.

This framework provides (i) a principled method for realizing trajectories from
flow-based Markov descriptions, (ii) a network-centric mechanism for studying
multi-step mobility structure induced solely by connectivity and timing, and,
most importantly, (iii) a conceptual foundation for inferring path-level
mobility information from temporally aggregated, privacy-preserving OD data.
In many applied settings, mobility data are released only after aggregation by
location and time slice, precluding direct observation of individual
trajectories. The results presented here demonstrate that, under explicit and
transparent structural assumptions, such aggregated data are nevertheless
compatible with a well-defined distribution over multi-step trajectories.

Viewed abstractly, the proposed generator defines a flexible,
multi-parameter family of probability distributions over trajectories, in
which structural modules (e.g., hubs, corridors, directional bias schedules)
may be included or omitted depending on what coarse information is available.
While no parameter estimation or data fitting is undertaken in this paper,
the construction provides a natural scaffold for future inference problems in
which aggregate OD data constrain the admissible trajectory distributions.
Crucially, this is not a multi-agent or behavioral model: trajectories are not
interpreted as decisions made by persistent individuals, but as realizations of a
prescribed, population-level stochastic dynamics in which individual identity
carries no modeling meaning.

A closely related motivation arises from the pseudo-Markov chain framework introduced in \cite{foster2025pseudomarkov}, where spatio-temporally aggregated OD data are used to define time-dependent transition matrices acting on population distributions. That formulation deliberately abstracts away from realized paths, modeling mobility entirely at the level of collective flow evolution. The present work complements this perspective by providing a concrete mechanism for realizing such prescribed, time-dependent Markov dynamics as synthetic trajectories—without introducing memory, individual heterogeneity, or behavioral assumptions. In this way, the generator developed here makes explicit the trajectory-level structure implicitly encoded in aggregate flow descriptions, while remaining fully aligned with a flow-first, privacy-preserving modeling philosophy.

At an aggregate level, the emphasis on externally driven flow dynamics parallels macroscopic traffic models, where mobility is treated as the evolution of densities and fluxes on a network rather than as the outcome of individual decisions \cite{lwr1955,richards1956,daganzo1995}. Unlike deterministic traffic-flow equations, however, our framework encodes these dynamics directly as a time-dependent Markov process and realizes them through stochastic, memoryless trajectories.

The Markov dynamics specified in this framework are not intended to be arbitrary. Rather, the time-dependent transition kernels encode coarse, external pressures acting on mobility at the network level. Gravity terms, center–periphery asymmetries, hub and corridor structure, and scheduled bias multipliers can be interpreted as simplified representations of recurring incentives and constraints—such as work schedules, land-use organization, transportation capacity, or policy-driven accessibility—that shape collective movement without resolving individual intent. From this perspective, the dynamics resemble a form of self-organization driven by network topology and temporally varying activity attractors (gravitational centers), in which mass redistributes coherently in response to shared conditions. Similar ideas appear implicitly in urban scaling and mobility studies that emphasize flow, coupling, and collective regularities over individual decision-making, as well as in broader work on stochastic transport on networks \cite{bettencourt2013}.

Related but distinct are large-scale transport and activity-based simulators, such as
SimMobility developed at MIT \cite{adnan2016simmobility}, SUMO \cite{krajzewicz2012sumo},
and MATSim-like frameworks \cite{horni2016matsim}.
 These systems generate synthetic trajectories by assigning agents activity schedules, behavioral rules, and route choices on detailed transportation networks, often calibrated to survey or census data. While stochastic at various stages, they remain agent-centric: agents possess persistent identities, heterogeneous preferences, and planned sequences of activities, and trajectories are the primary objects of interest. Our model differs fundamentally from this paradigm. We do not simulate activities, demand, or decision-making, and we do not attribute meaning to individual paths. Instead, trajectories arise as realizations of a prescribed, time-dependent Markov dynamics defined entirely by network structure and externally specified biases.

Privacy considerations fundamentally shape modern mobility research. While fine-grained spatial–temporal mobility data can enable trajectory reconstruction and re-identification \cite{dgbcd:opcumpd,xtlzfj:trfa}, many contemporary datasets therefore provide only aggregated origin–destination (OD) flows over fixed time intervals, discarding individual identifiers and temporal continuity to mitigate these risks \cite{dgbcd:opcumpd,xtlzfj:trfa}.
Such aggregate representations are also standard in large-scale, population-based mobility models, where movement is treated as flux between subpopulations rather than individual trajectories \cite{tizzoni2014od,cuebiqod,safegraphod}. In this work, we operate explicitly at this coarser level of abstraction, focusing on structural mobility patterns defined by network connectivity (e.g., corridors, hubs, and regional flows) rather than fine-grained traces; the aggregation scale assumed here deliberately excludes the spatial and temporal resolution required by known trajectory-recovery attacks \cite{xtlzfj:trfa}. At this level, we generate trajectories corresponding to Privacy-Enhanced Persons (PEPs)—
indistinguishable, memoryless movers that are interchangeable at any location where they meet, and that represent the maximal trajectory information recoverable after such
spatio-temporal aggregation, as formalized in \cite{foster2025pseudomarkov}.

The remainder of this paper is organized as follows.
Section~\ref{sec:genmod} presents the trajectory generation model in detail.  We first describe the spatial
discretization and construction of the undirected base graph, followed by the
selection of reference centers and the construction of hubs, corridors, feeder
paths, and metro links that define the overlay mobility network.
We then introduce node potentials and directed edge attributes that support
asymmetric, time-dependent dynamics on this network.

 Section~\ref{sec:pep-generation} formulates the population evolution and trajectory realization procedure.
 We define a family of time-dependent, column-stochastic Markov
matrices acting on population distributions, describe their construction from
network geometry and modular bias components, and explain how synthetic,
memoryless trajectories are realized for indistinguishable movers.
This section also introduces post hoc attribution of physical travel distance
and travel time to realized transitions, enabling time-elapsed mobility measures
without altering the underlying Markov dynamics.

Section~\ref{sec:verification} verifies the consistency of the trajectory realization with the
prescribed network-driven dynamics. We compare multi-step transition matrices
obtained by direct composition of the Markov kernels with those empirically
estimated from realized trajectories, and quantify agreement using matrix- and
column-level discrepancy measures.

Finally, the appendix states the form of the Perron--Frobenius theorem relevant to
the daily evolution matrix and justifies the existence and uniqueness of the
periodic invariant population distribution used throughout the paper.

Together, these sections establish a network-first methodology for realizing mobility trajectories from prescribed, time-dependent Markov dynamics and for validating their aggregate consistency without recourse to individual-level behavioral modeling.

 \section{The Trajectory Generation Model} \label{sec:genmod}
This section describes the construction of the network-driven mobility generator. The model separates cleanly into two conceptual layers: a static, spatially embedded mobility network encoding admissible routes and hierarchical structure, and a time-dependent Markov dynamics defined on that network. The former specifies where movement is possible and how locations are organized (hubs, corridors, feeders), while the latter prescribes how population mass redistributes over time through externally imposed, schedule-driven transition kernels. Synthetic trajectories arise as memoryless realizations of this dynamics and are discussed in Section \ref{sec:pep-generation}.

 \subsection{Spatial discretization and base graph}  \label{subsec:spacdiscbasegrph}
We begin by defining a cell space that is independent of any mobility data or behavioral assumptions. A geographic region of interest is specified by a polygonal boundary obtained from open geographic sources, optionally expanded by a fixed buffer distance to reduce boundary artifacts. This region defines the admissible domain within which all subsequent network construction takes place.

The region is discretized using the H3 hierarchical hexagonal grid
\cite{uberh3} at a fixed resolution.  Each H3 cell whose hexagonal polygon intersects the region boundary is retained as a node in the network, and is represented by the geographic centroid of the cell. This discretization yields a finite set of cells with uniform local geometry and bounded degree.

\begin{figure}[H]
  \centering
  \includegraphics[width=0.9\linewidth]{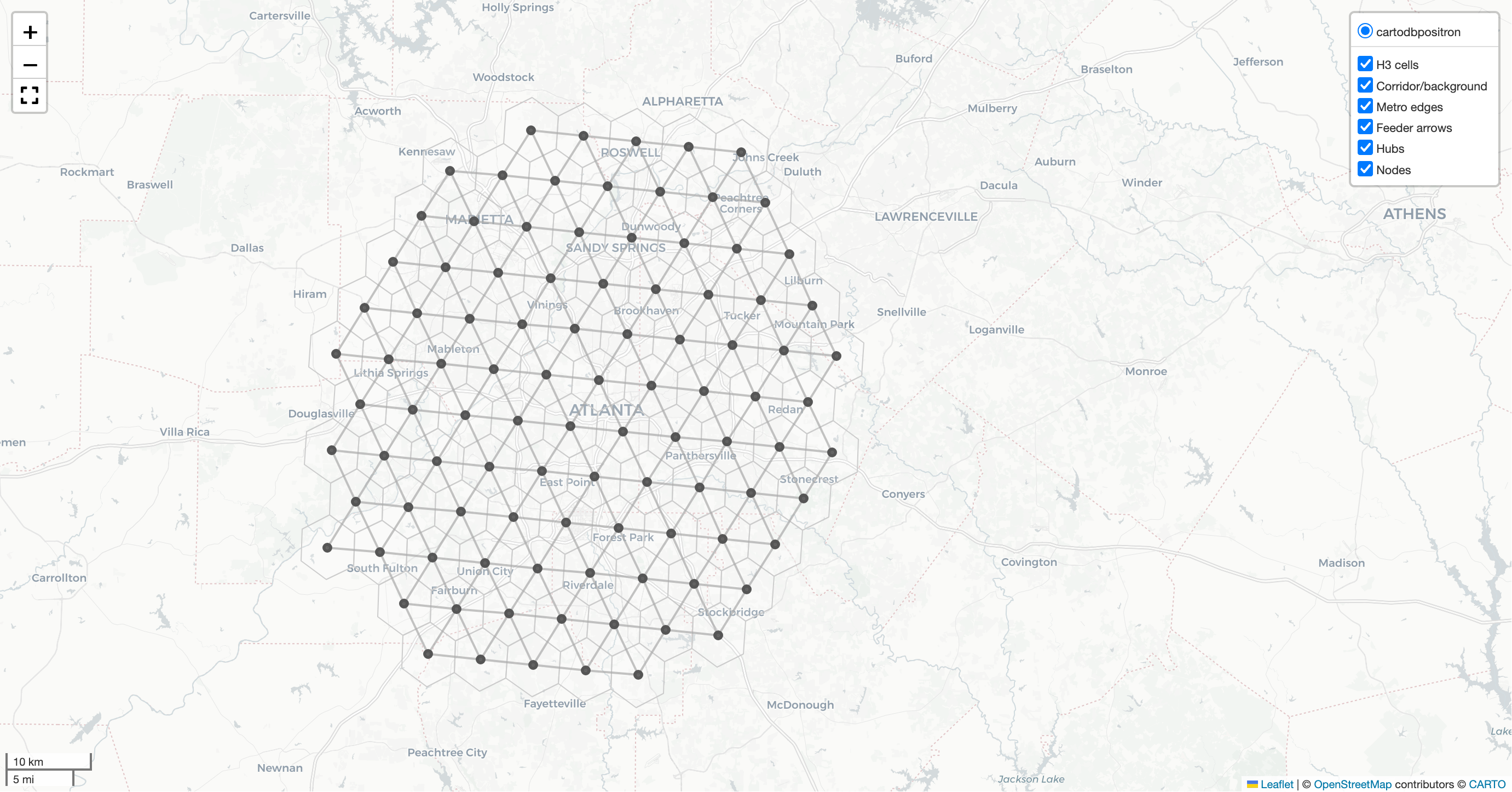}
  \caption{H3 spatial discretization and undirected base graph for the study region
  (Atlanta shown for geographic scale only). Hexagonal tiles indicate H3 cells at
  resolution~6, while nodes mark the corresponding cell centers.
  Adjacent cell centers are separated by approximately $6.44\,\mathrm{km}$ on average,
  and edges represent immediate H3 adjacency.}
  \label{fig:h3-base-grid}
\end{figure}

An undirected base graph is then constructed by connecting each H3 cell to its immediate
neighbors in the H3 grid. Two cells are adjacent if they are at H3 grid distance one,
corresponding to shared hexagonal boundaries. In all experiments reported here, we use
H3 resolution~6, for which the mean distance between centers of adjacent cells is
approximately $6.44\,\mathrm{km}$. This base graph encodes only physical proximity and
connectivity, and contains no directionality, hierarchy, or preference. At this stage,
the network is homogeneous and isotropic, serving as a topologically uniform substrate
upon which additional structural features are imposed in later steps
(Figure~\ref{fig:h3-base-grid}).

Although the figures in this paper use the Atlanta metropolitan area for visual reference,
the choice of region plays no role in the construction of the overlay network beyond setting
geographic scale and distances; all network structure, hierarchy, and dynamics are generated
independently of region-specific infrastructure or mobility data.

\subsection{Centers, hubs, and corridor structure} \label{subsec:hubs-corridors}

The purpose of this construction is not to model specific infrastructure, but to impose a controlled, hierarchical routing structure on an otherwise homogeneous spatial grid.
Starting from the undirected base graph, we introduce hierarchical structure by designating a small subset of nodes as hubs and organizing connectivity around them. A reference center is first selected to anchor the construction; this may be specified explicitly (e.g., a known city center) or determined geometrically from the spatial distribution of nodes. Distances from this center are then used to stratify the region into concentric radial bands, within which a prescribed number of hubs are selected from existing grid cells subject to minimum separation constraints. This procedure yields a sparse set of hubs distributed across space while respecting the underlying discretization and regional boundary.

Hubs are connected to the reference center by constructing a corridor backbone on the base graph. 
For each hub, we compute a shortest path to the center on the unweighted H3 adjacency graph using a standard shortest-path search, equivalent to Dijkstra’s algorithm with unit edge weights \cite{dijkstra1959}, and all edges along these paths are retained as backbone edges. 
This produces a tree-like structure rooted at the center, embedded within the base grid, along which long-range connectivity is concentrated. 
Paths may overlap and share segments naturally, allowing corridors to merge as they approach the center without enforcing explicit hub-to-hub links.

All non-hub nodes are connected to the hub backbone via feeder paths. 
For each node that is not designated as a hub, we identify the nearest hub in terms of graph distance on the base H3 adjacency. 
A shortest path from the node to that hub, computed using the same unweighted shortest-path procedure, is then added to the corridor structure, and the edges along this path are labeled as feeder edges. 
This ensures that every node in the region is connected to the backbone through at least one admissible route, producing a connected overlay network embedded within the base grid.

\begin{figure}[H]
  \centering
  \includegraphics[width=0.9\linewidth]{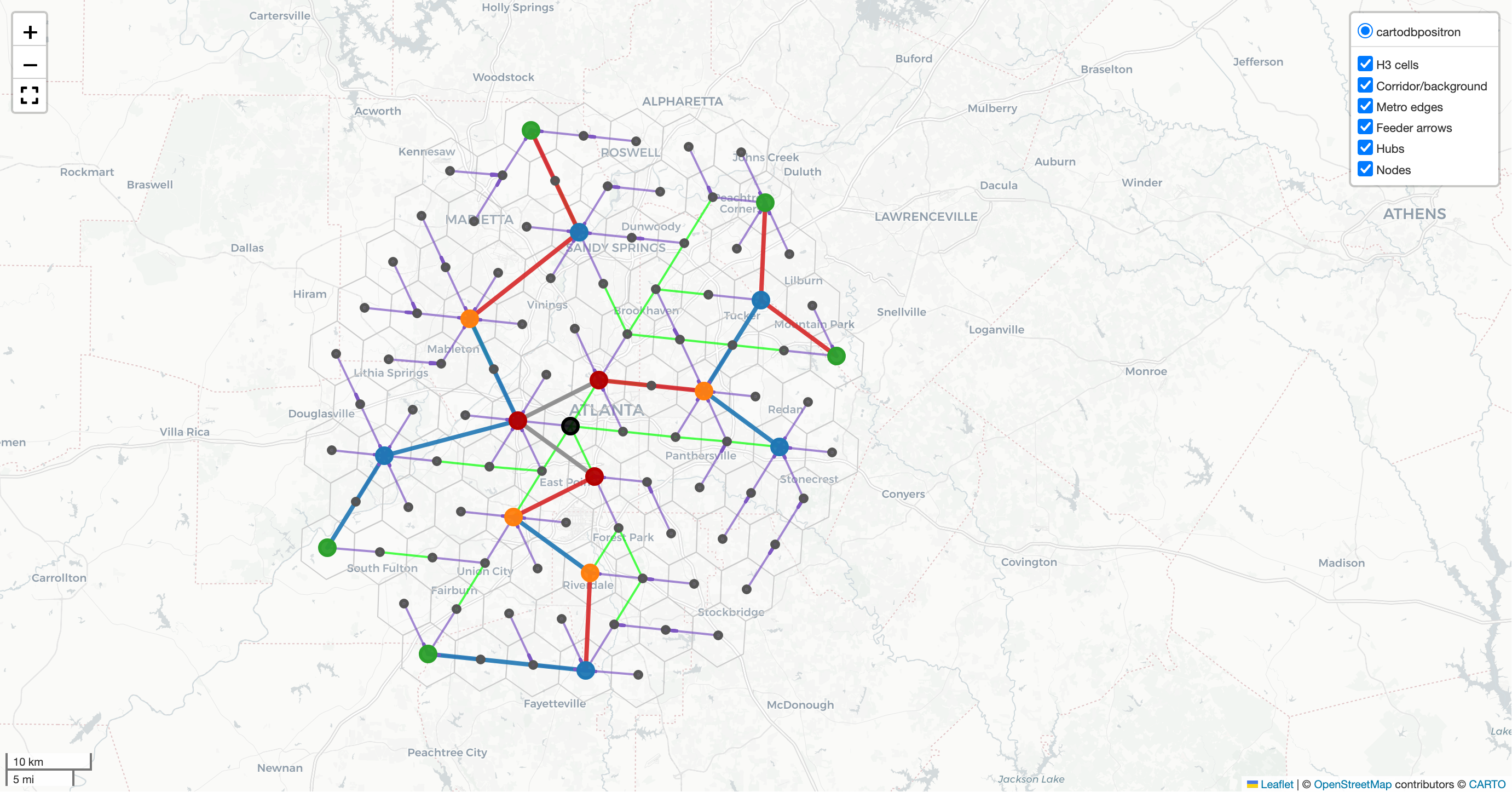}
\caption{Corridor overlay network constructed on top of the H3 base grid.
Nodes correspond to H3 cells, with the reference center shown as a black
node and hub locations shown as larger markers.
Highlighted edges define a hierarchical routing structure: green edges
indicate hub--backbone connections linking each hub to the central anchor,
while purple edges denote feeder paths connecting non-hub cells to hubs.
Where uniquely defined, feeder direction is indicated by darker terminal
segments. Metro links between hubs are drawn with increased line width.}
  \label{fig:h3-corridors}
\end{figure}

Finally, a sparse set of metro edges is constructed between hubs to represent high-capacity, long-range connections. Each hub is linked to a geometrically closer hub that lies nearer to the center, forming a directed tree over the hubs based on radial distance. These metro edges bypass the local grid adjacency and provide express connectivity across the region. Together, the hub backbone, feeder paths, and metro edges define an overlay graph that captures hierarchical organization, long-range corridors, and local access, while remaining fully embedded in the underlying spatial discretization (Figure~\ref{fig:h3-corridors}). All edges added to the overlay—whether backbone, feeder, or metro—are symmetrized after construction, so that each admissible route is traversable in both directions; this guarantees that the resulting directed mobility graph is strongly connected.

\subsubsection{Node potentials and directional orientation}
To support directionally biased dynamics in later stages of the model,
we associate a scalar potential with each node that encodes its position
within the hub--center hierarchy. Specifically, each node $i$ is assigned
a potential
\[
\Psi_i = d(i, c),
\]
where $d(i, c)$ denotes the shortest-path distance from node $i$ to the
reference center $c$ on the base H3 adjacency graph. Nodes closer to the
center therefore have lower potential values.

For each directed edge $(j \rightarrow i)$ in the overlay network, we
define a directional sign
\[
\sigma_{ij}
=
\operatorname{sign}\!\left(\Psi_j - \Psi_i\right)
\in \{-1, 0, +1\},
\]
where $\sigma_{ij}=+1$ indicates movement toward lower potential
(i.e., inward toward the center), $\sigma_{ij}=-1$ indicates movement
toward higher potential (outward from the center), and $\sigma_{ij}=0$
corresponds to neutral edges with no net potential change. This sign is
fixed by the network geometry and hierarchy, and does not depend on time.

The directional sign $\sigma_{ij}$ is used in the construction of the
time-dependent transition matrices to apply scheduled directional bias
factors, enabling asymmetric flow along corridors and feeders while
preserving a Markovian  dynamics.

\section{Population Evolution and Trajectory Realization}
\label{sec:pep-generation}
In this section we specify the time-dependent Markov dynamics that govern population evolution on the network constructed in Section~\ref{sec:genmod}, and describe how synthetic trajectories are realized from these dynamics. The key modeling choice is to prescribe a family of column-stochastic transition matrices directly from network geometry, distance decay, and externally imposed temporal biases, rather than estimating them from trajectory data. Population evolution is treated deterministically at the distribution level, while trajectories arise as independent, memoryless realizations of the same dynamics.

We adopt a matrix-based formulation in which transitions are given by a time-indexed family of Markov matrices acting on population distributions over spatial locations. In this section we describe how this formalism is instantiated on the network constructed in Section~\ref{subsec:spacdiscbasegrph}, and how synthetic trajectories are realized from the resulting dynamics.

\subsection{Time-dependent Markov matrices}

Let $V=\{1,\dots,N\}$ denote the set of spatial locations (H3 cells). For each discrete time-step $t$, we define a column-stochastic matrix
\[
M_t \in \mathbb{R}^{N\times N},
\qquad
\sum_{i=1}^N (M_t)_{ij} = 1 \quad \forall j \in V,
\]
where $(M_t)_{ij}$ gives the probability of transitioning from location $j$ to location $i$ during time-step $t$.

The matrix $M_t$ is not inferred from trajectory data. Instead, it is specified exogenously as a function of:
\begin{itemize}
  \item the underlying spatial adjacency graph,
  \item the corridor and hub overlay described in Section~\ref{subsec:hubs-corridors},
  \item geometric distance between locations,
  \item and time-dependent weighting modules encoding center--periphery asymmetries, directional biases, and corridor preferences.
\end{itemize}

Self-loops $(i=j)$ are explicitly included and represent persistence at the same location during the current time-step. 

A deliberate feature of this construction is that transition probabilities are normalized independently for each origin, so that any origin-specific mass factors cancel identically; only relative destination attractiveness and network structure influence movement decisions.

\subsection{Population evolution}

Let $\mathbf{p}_t \in \mathbb{R}^N$ denote the population mass over locations at time-step $t$, with
\[
\mathbf{p}_t \ge 0,
\qquad
\sum_{i=1}^N (\mathbf{p}_t)_i = 1 .
\]
Following Section~4 of~\cite{foster2025pseudomarkov}, the population evolves deterministically according to
\[
\mathbf{p}_{t+1} = M_t \, \mathbf{p}_t .
\]
An initial distribution $\mathbf{p}_0$ is specified externally. In the implementation, $\mathbf{p}_0$ may be chosen uniformly or as a fixed point of the composed daily matrix, but the construction does not depend on this choice.
This evolution equation defines a time-inhomogeneous Markov process on the space of population distributions. No individual-level information is introduced at this stage.

\subsubsection{Gravity-type interaction kernel.}

Transitions between locations are constructed from a gravity-type interaction kernel.
In its classical form, a gravity model assigns an unnormalized flow intensity
\[
F_{ij}
=
k \, \frac{m_i \, m_j}{d_{ij}^{\beta}},
\]
where $m_i$ and $m_j$ are destination and origin masses, $d_{ij}$ is the distance between locations,
$\beta>0$ is a distance-decay exponent, and $k$ is a normalization constant
\cite{bhg:slht,simini2012,kwon2023multiple}.

In the present framework, we adopt this structure as a \emph{time-dependent interaction kernel}
rather than as a standalone flow model. The masses and normalization are allowed to vary with time,
and the kernel is embedded within a Markov transition matrix defined on a constrained mobility
network.

Specifically, each location $i$ is assigned a time-dependent mass $m_i(t)$ encoding
coarse spatial asymmetries such as center--periphery structure, hub status, and externally imposed
temporal schedules. Origin mass factors $m_j(t)$ are included at the kernel level but cancel under
Markov normalization, as described below. Distances $d_{ij}$ are computed from the spatial embedding
of the H3 cells and remain time-independent.

\subsubsection{Structural edge weights.}
The factor $\omega_{ij}$ is a static, nonnegative weight encoding the structural connectivity of the overlay network. It reflects whether the edge $(j,i)$ lies on:
\begin{itemize}
  \item a hub backbone corridor,
  \item a feeder path,
  \item or a long-range metro connection between hubs.
\end{itemize}
Edges absent from the overlay have $\omega_{ij}=0$ and are therefore not admissible transitions.

\subsubsection{Time-dependent bias factors.}
The factor $\phi_{ij}(t)$ aggregates all remaining time-dependent modifiers that act on admissible edges. These include:
\begin{itemize}
  \item feeder-specific scaling factors that distinguish primary and secondary attachments,
  \item hub-related scaling applied to outgoing metro edges,
  \item and directional biases that favor motion toward or away from designated reference centers.
\end{itemize}
Each component may follow its own externally specified schedule, evaluated as a function of the current time-step $t$. These schedules are independent across modules and are applied multiplicatively.

\subsection{Construction of the transition matrices}

We now give the explicit construction of the matrices $M_t$ introduced above.
The mobility dynamics are specified by a family of time-dependent,
column-stochastic Markov matrices $\{M_t\}_{t\in\mathcal{T}}$ acting on
population vectors. For each origin node $j$, the diagonal entry
$(M_t)_{jj}$ represents the probability of remaining at $j$ during time-step $t$.

\[
(M_t)_{jj} = s_j(t),
\]
where $s_j(t)\in[0,1]$ is a time-dependent stay probability.

For $i \neq j$, transitions are defined by normalized edge weights.
Let
\[
w_{ij}(t)
=
k(t)\,
\frac{ {m_i(t)}^\alpha \, {m_j(t)}^\alpha }{ d_{ij}^{\beta} }\,
\omega_{ij}\,
\phi_{ij}(t),
\qquad i \neq j .
\]

Here $m_i(t)$ and $m_j(t)$ denote time-dependent destination and origin mass factors,
$d_{ij}$ is the geographic distance between locations $j$ and $i$, $\alpha$ is a mass-scaling parameter, $\beta>0$ is a
distance-decay exponent, $\omega_{ij}$ encodes static structural connectivity of the
overlay network, and $\phi_{ij}(t)$ aggregates all remaining time-dependent directional
and structural bias factors. The scalar $k(t)$ is a time-dependent normalization constant
introduced to ensure compatibility with the Markov normalization described below.

Although the interaction kernel includes both origin and mass factors,
the origin mass $m_j(t)$ does not influence the resulting transition probabilities.
As shown below, Markov normalization is performed independently for each origin $j$,
and the factor $m_j(t)$ cancels identically from all outgoing transitions from $j$.
As a result, the effective transition structure depends only on relative destination
attractiveness, distance decay, network connectivity, and time-dependent biases,
and not on the absolute mass assigned to the origin.

\subsubsection{Center--periphery structure and flow objectives}

The purpose of the modular weighting scheme is to generate structured,
time-dependent flows between a central urban core and a surrounding
periphery. To this end, we designate a small set of spatial cells near a
geographic or network-defined reference point as the \emph{center}.
All remaining locations are treated as peripheral in this deliberately
simplified representation of urban structure.

Each node is assigned a scalar potential measuring its graph distance to
the center, which induces a consistent notion of inward and outward
movement along the overlay network. The transition weights are then
constructed so that, during selected periods of the day, movement toward
the center is preferentially amplified, while during other periods,
movement toward the periphery is favored.

The resulting dynamics are not inferred from data. Instead, weights,
biases, and schedules are chosen explicitly to produce canonical
diurnal patterns—such as morning inflow and evening outflow—at the level
of aggregate network structure.

The transition matrices are assembled from a collection of modular components, not all of which are required in every configuration; together, they provide a flexible vocabulary for encoding structural and temporal asymmetries in the flow dynamics.

\begin{itemize}

\item \textbf{Module A (stay probability schedules).}
Specifies the time-dependent stay probability $s_j(t)$ at each location $j$.

\item \textbf{Module B (node mass schedules).}
Defines time-dependent  masses $m_i(t)$ for each node.

\item \textbf{Module C (hub stay adjustment).}
Applies additional time-dependent scaling to the stay probabilities of hub nodes.

\item \textbf{Module D (hub mass modulation).}
Applies multiplicative factors to the mass schedules of hub nodes.

\item \textbf{Module E (feeder and corridor scaling).}
Assigns static structural weights distinguishing feeder paths, corridor backbones,
and local connections.

\item \textbf{Module F (metro link schedules).}
Applies time-dependent scaling to long-range metro edges between hubs.

\item \textbf{Module G (directional bias schedules).}
Applies time-dependent directional bias factors to edges based on their orientation
relative to the reference center.

\end{itemize}

As a representative example, we detail the construction of Module~G (directional bias); other modules follow the same scheduling pattern.
\subsubsection{Module G: Directional bias schedules}

Module~G introduces time-dependent directional asymmetries in movement
along network edges. Each directed edge \((j \rightarrow i)\) carries a
fixed directional sign \(\sigma_{ij}\in\{-1,0,+1\}\), defined by the
network construction in Section~\ref{subsec:hubs-corridors} and encoding
whether the edge is oriented outward from the reference center (\(-1\)),
inward toward the center (\(+1\)), or neutral (\(0\)). 
Module~G does not alter this geometric orientation; instead, it applies
time-dependent bias factors to edges based on their preassigned
directional sign, thereby inducing asymmetric flow patterns that vary
over the daily cycle.

The directional bias factor $\phi_{ij}(t)$ appearing above is defined as
\[
\phi_{ij}(t)
=
b_{\sigma_{ij}}(t),
\]
where $b_{+1}(t)$, $b_{-1}(t)$, and $b_{0}(t)$ are scalar, time-dependent
multipliers associated with inward, outward, and neutral directions,
respectively.

Each multiplier is specified by a schedule over the day. A schedule
consists of a sequence of time-steps
\[
(\tau_s,\tau_e,v),
\]
meaning that the multiplier ramps linearly from its current value at
time $\tau_s$ to the value $v$ at time $\tau_e$, and is held constant
until the next scheduled ramp. Outside all specified intervals, the
multiplier remains at its most recently attained value.

\subsubsection{Example.}
Consider an inward directional bias $b_{+1}(t)$ with the schedule
\[
(06{:}00,\,11{:}00,\,5.0),
\qquad
(15{:}00,\,20{:}00,\,1.0).
\]
Starting from a baseline value of $1.0$, the inward bias increases
linearly between 06{:}00 and 11{:}00, reaching $5.0$ at 11{:}00, and is
held at that level until 15{:}00. Between 15{:}00 and 20{:}00 it ramps
back down to $1.0$, where it remains for the rest of the day. During the
morning peak, inward-directed edges are therefore strongly favored,
while in the evening the directional preference is neutralized.

This mechanism allows the transition matrices $M_t$ to encode recurring
collective patterns such as morning inflow toward central areas and
evening outflow, without introducing agent memory or individual intent.
Directional bias schedules act multiplicatively on edge weights and are
fully compatible with the Markov structure of the dynamics.

 \subsubsection{Normalization}

For each origin $j$, the off-diagonal weights $\{w_{ij}(t)\}_{i \neq j}$
are normalized so that their sum equals $1 - s_j(t)$:
\[
(M_t)_{ij}
=
(1 - s_j(t))\,
\frac{ w_{ij}(t) }{ \sum_{i' \neq j} w_{i'j}(t) },
\qquad i \neq j .
\]

Because the origin mass $m_j(t)$ appears multiplicatively in every
$w_{ij}(t)$ for fixed $j$, it cancels under normalization. As a result,
the transition probabilities depend on destination masses, distance,
network structure, and time-dependent biases, but not on the absolute
mass assigned to the origin.

\subsection{Trajectory realization}

Synthetic trajectories are realized by interpreting the population dynamics above as a flow matrix, instantiating memoryless movers that are interchangeable when they share the same location at a time-step. Each mover occupies a location $j$ at time-step $t$ and performs a single categorical (single-outcome) draw
 from the distribution defined by column $(M_t)_{\cdot j}$, i.e., hops according to the column transition probabilities to one of the destinations. The resulting destination determines the carrier's location at time-step $t+1$.
This procedure is repeated independently for all movers and all time-steps.  Aggregate OD counts emerge as empirical sums of these categorical transitions.

\subsection{Choice of the initial population distribution} 
\label{subsec:initial-population}
Because the Markov dynamics are time-inhomogeneous, the notion of a stationary distribution is replaced by a periodic invariant distribution over one full daily cycle.
The evolution of population mass in the model is governed by the time-dependent Markov matrices
\(\{M_t\}_{t=1}^{T}\), where one full day corresponds to a fixed period of \(T\) discrete time steps.
Over one period, these matrices define a daily evolution matrix
\[
\mathbf{p}_{T} = Q\,\mathbf{p}_0,
\qquad
Q := M_T M_{T-1} \cdots M_1,
\]
where \(\mathbf{p}_t\) denotes the population distribution over nodes at time \(t\).

By construction, each \(M_t\) is column-stochastic, and hence so is the daily matrix \(Q\).
Moreover, the overlay mobility network is connected, with hubs, corridors, feeders, and metro
edges ensuring that every node can reach every other node within a single daily cycle.
As a consequence, the matrix \(Q\) is irreducible.

Under these conditions, the Perron--Frobenius theorem for nonnegative matrices guarantees the
existence of a unique invariant population distribution \(\mathbf{p}^\ast\) satisfying
\[
Q\,\mathbf{p}^\ast = \mathbf{p}^\ast,
\]
up to normalization (see Appendix~\ref{app:pf}).

In this work, we choose the initial population distribution \(\mathbf{p}_0\) to be this periodic
fixed point \(\mathbf{p}^\ast\). This choice eliminates transient effects associated with arbitrary
initialization and ensures that the realized trajectories reflect the intrinsic structure of the
prescribed time-dependent transition matrices rather than relaxation toward equilibrium.

\begin{figure}[H]
  \centering
  \includegraphics[width=0.9\linewidth]{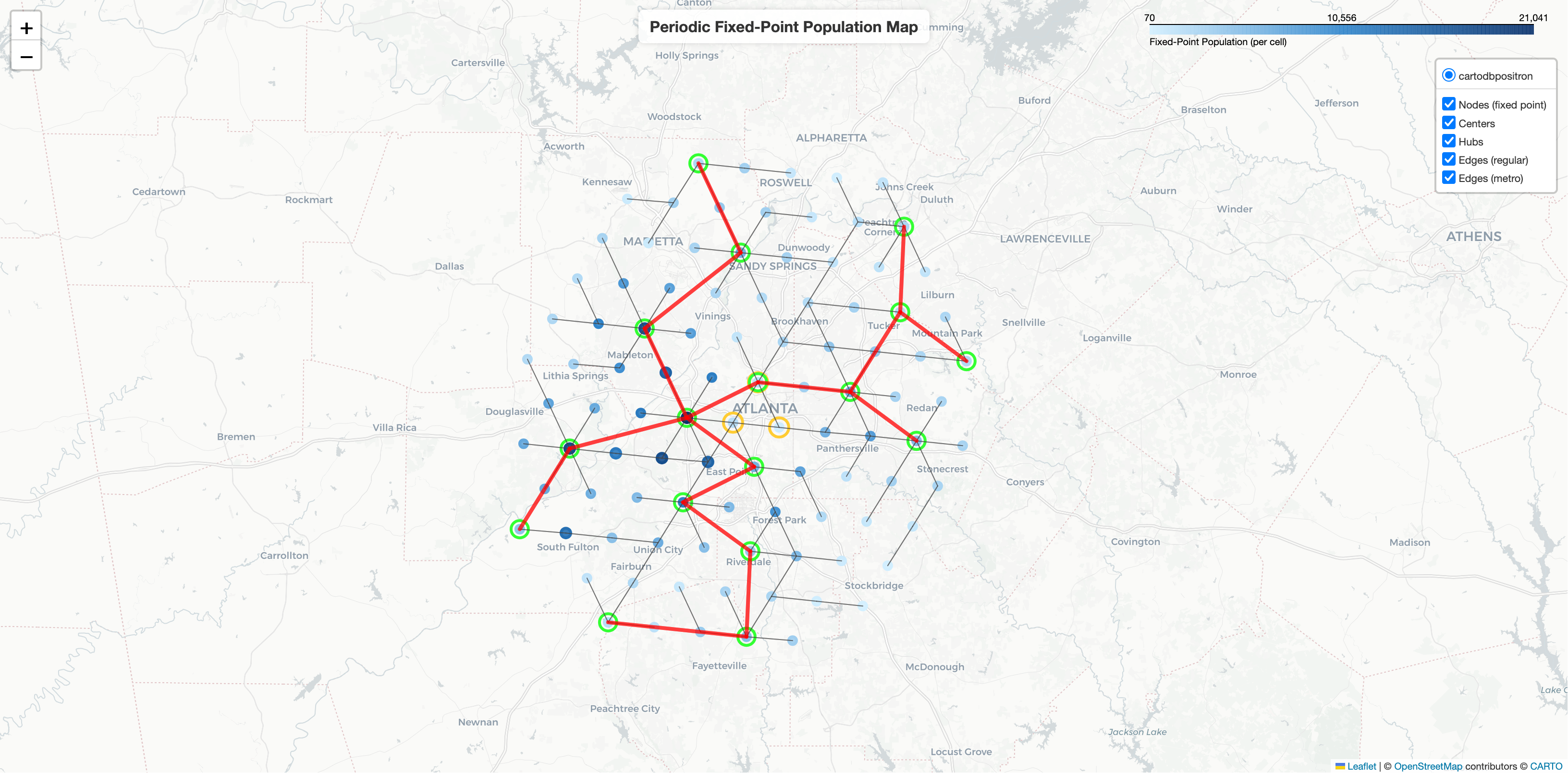}
\caption{Periodic fixed point population distribution \(\mathbf{p}^\ast\) of the daily
Markov matrix \(Q = M_T \cdots M_1\).
Node color intensity represents the invariant population mass associated with each H3 cell,
with darker colors indicating higher population as shown in the colorbar.
This distribution is uniquely determined (up to normalization) by the 
OD transition probabilities encoded in the matrices \(\{M_t\}\), and reflects
the intrinsic spatial organization implied by the network-driven dynamics rather than
any imposed initial condition.
The Atlanta region is shown for geographic scale only.}
  \label{fig:periodic-fixed-point}
\end{figure}

Figure~\ref{fig:periodic-fixed-point} illustrates that the invariant population
distribution exhibits strong spatial structure induced entirely by network
connectivity, distance effects, and scheduled directional biases, despite the
absence of any individual-level behavioral assumptions.

\subsubsection{Remark (Significance of the periodic fixed point).}
The existence of the periodic fixed point \(\mathbf{p}^\ast\) has a strong and useful interpretation.
It implies that the long-run distribution of population over locations is fully determined up to normalization by
the collection of  OD transition probabilities encoded in the matrices
\(\{M_t\}\). In particular, knowledge of how mass is redistributed from each origin to all possible
destinations within each time-step is sufficient to recover a stable, day-periodic population
distribution, without reference to individual trajectories, activity patterns, or behavioral
heterogeneity.

This highlights the extent to which aggregate flow structure alone constrains macroscopic mobility
outcomes. The fixed point \(\mathbf{p}^\ast\) represents the unique population configuration that is
self-consistent with the prescribed network-driven transition dynamics, and may therefore be
interpreted as the intrinsic spatial organization implied by the observed (or specified) OD flows.

\subsection{Travel distance attribution}

The Markov transition matrices specify only the probability of movement between
spatial cells and do not encode physical travel distances. To attach a geometric
length scale to realized trajectories, we assign a travel distance to each
realized transition after destination selection.

Let \(j \rightarrow i\) denote a realized transition between two H3 cells at a
fixed resolution. For each ordered pair \((j,i)\), we precompute a distance
envelope \([d^{-}_{ji}, d^{+}_{ji}]\) based solely on the H3 geometry and the
geographic locations of the cell centroids. Let \(d_{ji}\) denote the great-circle
distance between the centroids of cells \(j\) and \(i\), and let \(D_{\text{hex}}\)
denote the diameter of an H3 hexagon at the chosen resolution.

For transitions between distinct cells (\(j \neq i\)), the envelope is defined as
\[
d^{-}_{ji} = \max\{0,\, d_{ji} - D_{\text{hex}}\}, 
\qquad
d^{+}_{ji} = d_{ji} + D_{\text{hex}}.
\]
This interval captures the range of plausible travel distances between points
lying anywhere within the two hexagonal cells.

For self-transitions (\(j=i\)), representing local movement within a single H3
cell, the envelope is defined as
\[
d^{-}_{jj} = 0,
\qquad
d^{+}_{jj} = D_{\text{hex}}.
\]

Conditional on a realized transition \(j \rightarrow i\), the travel distance is
then sampled independently from a uniform distribution on
\([d^{-}_{ji}, d^{+}_{ji}]\). This sampling is performed independently for each
trajectory and each time step.

Distance attribution is strictly post hoc: it does not influence the transition
probabilities, the construction of the Markov matrices, or the evolution of
population mass. Its sole purpose is to attach physically meaningful length
scales to realized trajectories, enabling downstream analysis of time-elapsed
flows, total distance traveled, and other distance-based mobility measures while
preserving a clear separation between probabilistic routing and geometric
interpretation.

\subsection{Travel time attribution}

For each realized transition \(j \rightarrow i\), a travel time is assigned
independently based on the realized travel distance and a randomly sampled
travel speed. Each PEP is assumed to move at a constant speed over the duration
of a single hop, with the speed drawn independently from a fixed distribution
representing plausible urban travel conditions (e.g., a truncated distribution over plausible urban commuting speeds). The travel time is then
computed as the ratio of the realized distance to the sampled speed and is
constrained to lie within the duration of the corresponding time-step. If the raw distance–speed ratio exceeds the time-bin duration, the travel time is truncated at the time-step length.

As with distance attribution, travel time assignment does not influence
transition probabilities or destination choice. The stochastic mobility
dynamics are governed entirely by the time-dependent Markov matrices, while
travel times serve only to attach physically interpretable temporal metadata to
each realized hop. At the aggregate level, origin--destination travel times are
reported by averaging over realized PEP hops within each
origin--destination--time-step, yielding consistent OD-level travel-time
summaries without altering the underlying flow structure.

\subsection{Generated data}

The trajectory generator produces both raw synthetic trajectories and
aggregated origin--destination (OD) data products directly from the same
underlying Markov dynamics. These outputs are generated simultaneously during
trajectory realization and are therefore internally consistent by construction.

\subsubsection{Trajectory-level outputs}

At the finest resolution, the generator produces raw PEP trajectories.
Each trajectory consists of a sequence of spatial locations indexed by discrete
time-steps, together with post hoc attributes attached to each realized hop.
Each PEP is assigned a persistent identifier that is used solely to link
successive transitions over time.

For every transition \(j \rightarrow i\), the output records:
\begin{itemize}
  \item the persistent PEP identifier,
  \item the origin and destination node identifiers,
  \item the time-step index,
  \item the realized travel distance,
  \item and the realized travel time.
\end{itemize}

These trajectories represent memoryless realizations of the prescribed
time-dependent Markov dynamics. The persistent identifiers function solely as bookkeeping labels: conditional on a location and time-step, realized hops are exchangeable across PEPs, and the identifiers carry no individual-level state, memory, or behavioral meaning.

\subsubsection{Aggregated OD outputs}

In parallel with trajectory generation, the code produces aggregated
origin--destination summaries over user-specified temporal windows.
For each OD pair \((j,i)\) and time-step, the following
quantities are computed directly from realized PEP hops:
\begin{itemize}
  \item total trip count,
  \item mean travel distance,
  \item median travel distance,
  \item standard deviation of travel distance,
  \item mean travel time,
  \item median travel time,
  \item and standard deviation of travel time.
\end{itemize}

These summaries are obtained by simple aggregation of realized transitions and
do not involve any additional modeling assumptions. Because both trajectory-level
and OD-level data are derived from the same realizations, the aggregated outputs
are guaranteed to be consistent with the underlying Markov matrices up to
finite-population sampling effects.

Although all examples in this paper use the H3 hexagonal grid, the construction
does not depend on any specific spatial discretization. The same procedure applies
to alternative tessellations such as Geohash or square grids, provided that:
(i) adjacency relationships are defined,
(ii) cell centroids are available for distance calculation, and
(iii) a bounded cell diameter can be specified for distance and time attribution.
The choice of discretization affects geometric resolution but does not alter the
Markov formulation, trajectory realization, or verification framework.

\begin{figure}[H]
  \centering
  \includegraphics[width=0.9\linewidth]{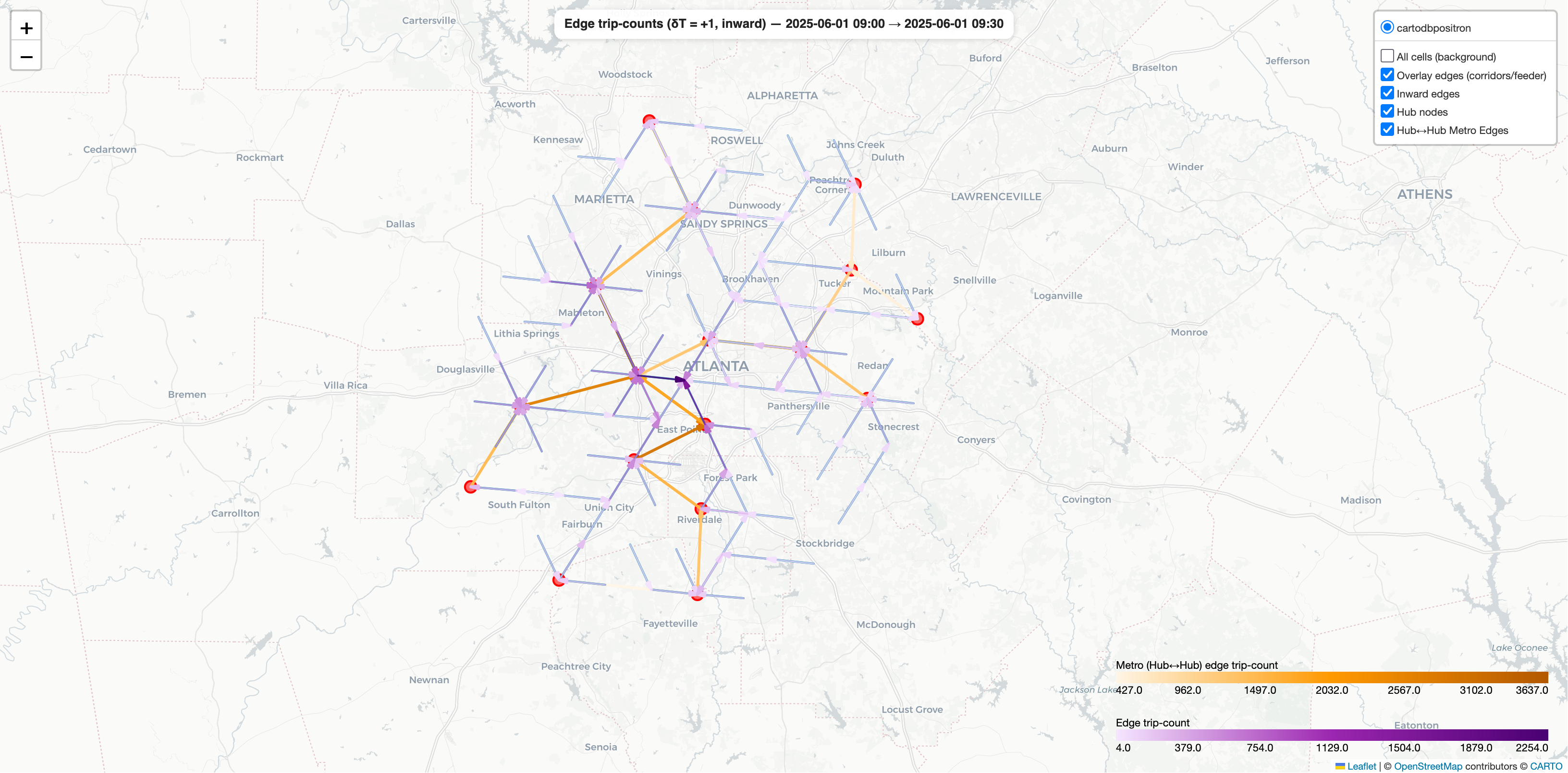}
\caption{Single time-step OD flows during the 09{:}00--09{:}30 time-interval, restricted to edges oriented inward toward the reference center.
Edge color intensity represents the magnitude of flow along each directed edge,
with darker colors indicating higher flow as shown in the colorbar.
Flows reflect the directional structure encoded in the transition matrix
\(M_t\) for this time step.}
  \label{fig:flow-inward}
\end{figure}

\begin{figure}[H]
  \centering
  \includegraphics[width=0.9\linewidth]{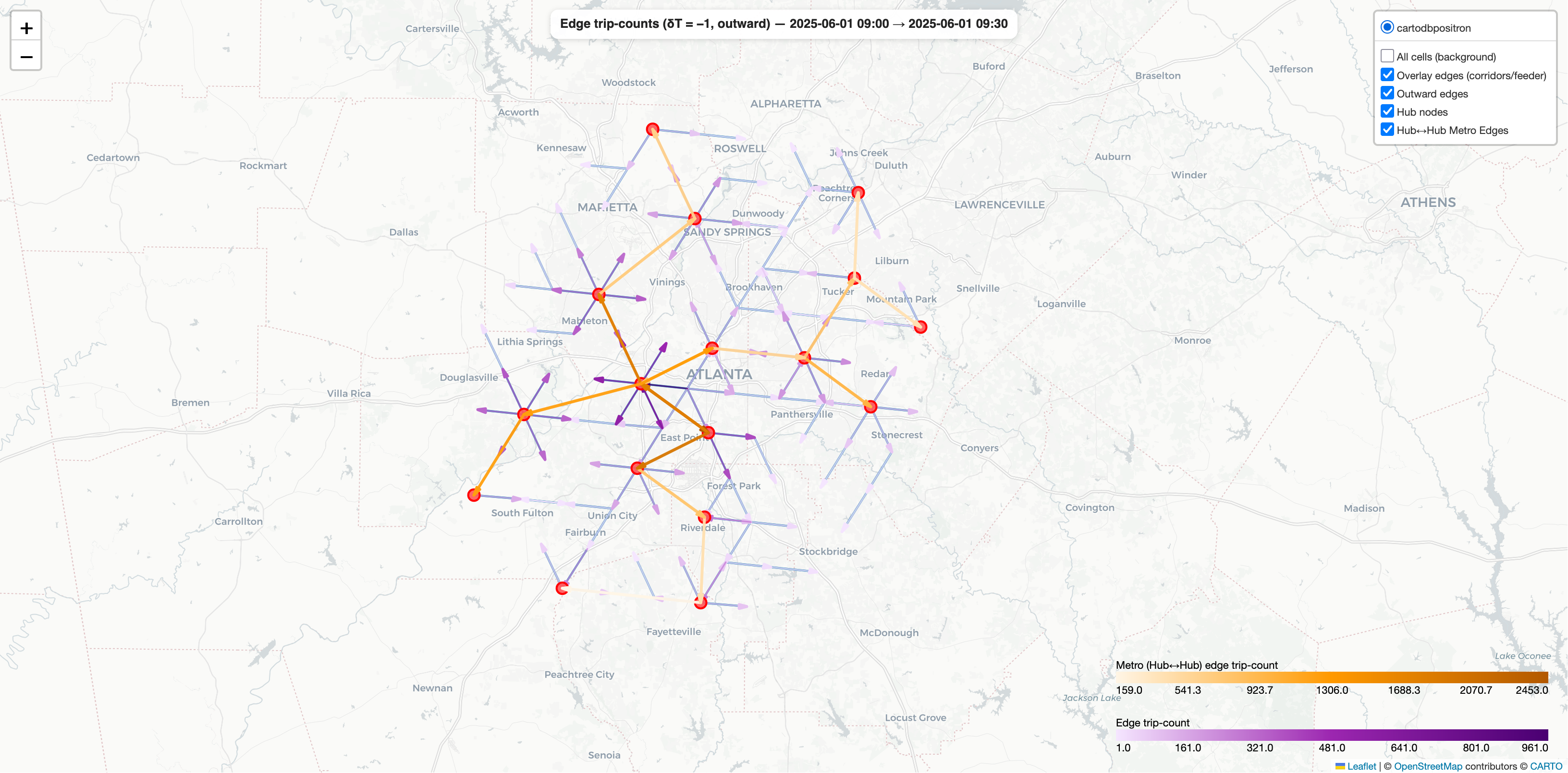}
\caption{Single time-step OD flows during the same 09{:}00--09{:}30
time-interval as Figure~\ref{fig:flow-inward}, restricted to edges oriented outward
from the reference center.
Edge color intensity represents the magnitude of flow along each directed edge,
with darker colors indicating higher flow as shown in the colorbar.}
  \label{fig:flow-outward}
\end{figure}

To illustrate the structure of the generated OD data at a fixed time,
Figures~\ref{fig:flow-inward} and~\ref{fig:flow-outward} show the realized
OD flows for a single representative time-step (09{:}00--09{:}30). Both figures
correspond to the same transition matrix \(M_t\), with the system initialized
at the periodic fixed point \(\mathbf{p}^\ast\) at 00{:}00 and evolved forward
to the 09{:}00--09{:}30 time-step from which the OD flows are collected. They
differ only in that flows are restricted to edges oriented inward toward the
reference center (Figure~\ref{fig:flow-inward}) or outward away from it
(Figure~\ref{fig:flow-outward}). Together, they provide a directional
decomposition of the OD traffic generated in a single time step. In all maps,
color intensity increases monotonically with population or flow magnitude.

\begin{figure}[H]
  \centering
  \includegraphics[width=0.9\linewidth]{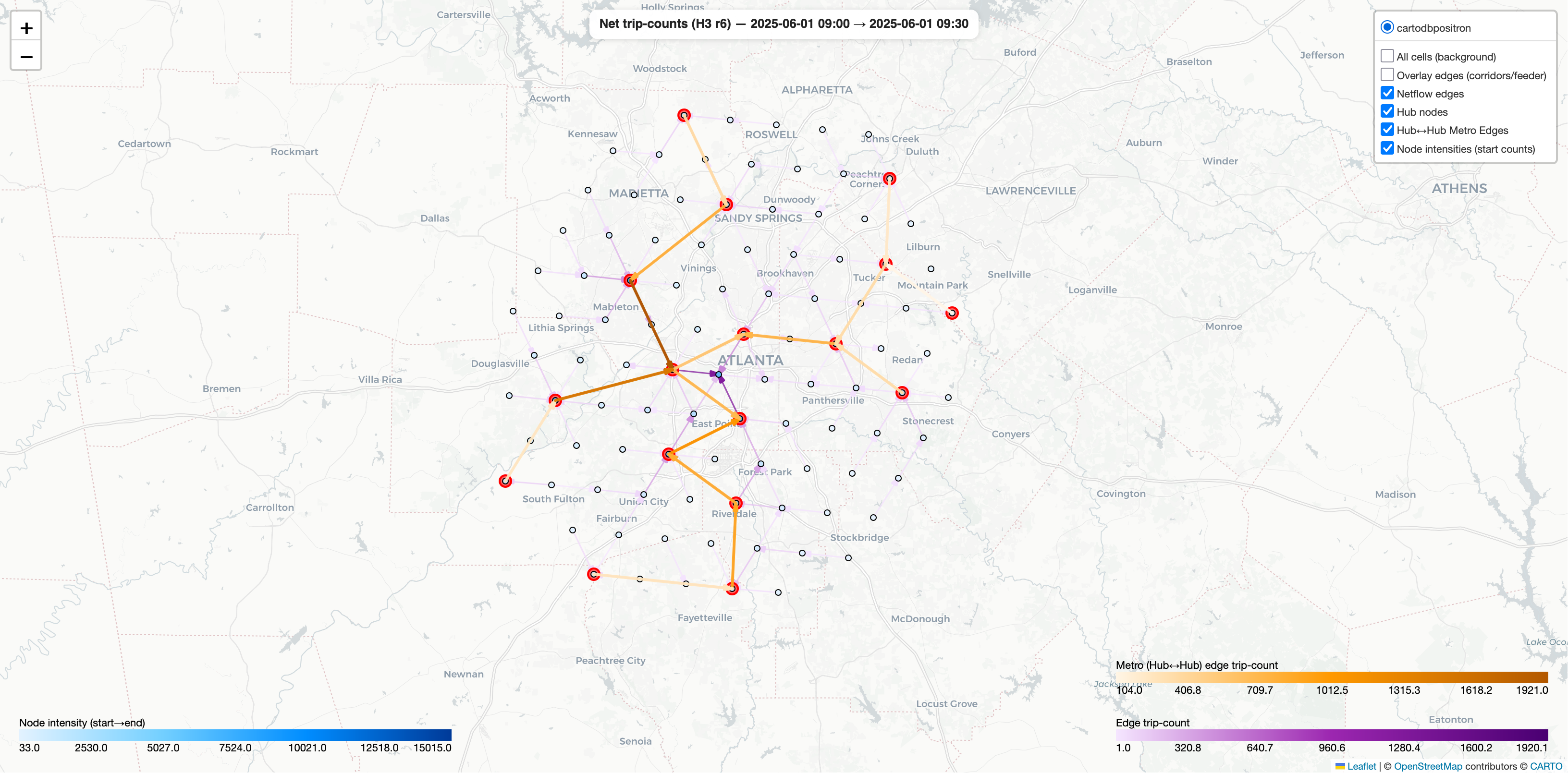}
\caption{Net OD flows during the 09{:}00--09{:}30 time-interval.
Edge color indicates the signed net flow magnitude, defined as inward flow minus
outward flow along each edge.
Higher net movements correspond to darker colors.
The predominance of inward net flows, particularly on edges close to the center,
is consistent with expected morning commuter dynamics and confirms that the 
configured directional biases produce the intended aggregate behavior.}
  \label{fig:flow-net}
\end{figure}

To further summarize the directional balance of movement at this time,
Figure~\ref{fig:flow-net} shows the corresponding \emph{net flow}  for the
same 09{:}00--09{:}30 time-step. Net flow on each edge is defined as the
difference between inward- and outward-oriented OD traffic along that edge.
Positive net flow therefore indicates dominant movement toward the reference
center, while negative values indicate dominant outward movement.

\subsection{Privacy interpretation}

Because the transition matrices $M_t$ are specified independently of any observed individual trajectories, and because realized paths are generated from interchangeable, memoryless movers
(exchangeable conditional on their current location), the resulting data contain
no linkable personal information. The generator corresponds to the limit in which all individual identity has been removed and only  network-constrained flow information is retained.

\section{Internal consistency between prescribed dynamics and trajectory realizations}
\label{sec:verification}

The purpose of this section is to verify internal consistency between the
prescribed time-dependent Markov dynamics and the realized PEP trajectories.
Synthetic trajectories are generated as memoryless realizations of the
single time-step transition matrices $\{M_t\}$ described in
Sections~\ref{sec:genmod} and~\ref{sec:pep-generation}.
From these trajectories we independently estimate an empirical end-to-end
transition matrix over a fixed time interval.
A necessary consistency requirement is that this empirical multi-step
transition matrix agrees, up to finite-sample effects, with the
multi-step transition matrix obtained by composing the prescribed
single time-step matrices $\{M_t\}$ over the same interval.
The analysis below is therefore a consistency check between OD-level and trajectory-level representations of the same underlying mobility dynamics, not an empirical validation against external data.

We verify this consistency by comparing two end-to-end transition matrices over a fixed temporal window:
one obtained by composing the prescribed single time-step transition matrices $\{ M_t \}$,
and one empirically estimated from realized PEP trajectories over the same interval.

\subsection{Multi-step transition matrices}

Consider a verification interval consisting of $m$ consecutive time-steps
$t_0, t_1,\dots, t_m$. The corresponding multi-step transition matrix implied by the model is
\[
A^{\text{prod}} = M_{t_m} M_{t_{m-1}} \cdots M_{t_1}
\]
which maps an initial population distribution $x_{t_0}$ at the start of the interval to the 
distribution at its end.

Independently, we construct an empirical end-to-end matrix from realized PEP trajectories.
Let $F_{ij}$ denote the number of PEPs whose first observed location in the interval is node $j$
and whose last observed location is node $i$. Let
\[
x_{t_0}(j) = \sum_i F_{ij}
\]
denote the number of trajectories originating at node $j$. The empirical transition matrix is
then defined column-wise as\footnote{In the verification experiments reported here, the number of realized
PEP trajectories \(K\) is sufficiently large that every origin node has
\(x_{t_0}(j)>0\) over the verification window, and this fallback case is therefore
never activated in practice. The identity-column convention is included for formal
completeness and robustness in finite-sample or sparse regimes, where some origin
nodes may contribute zero observed trajectories at the start of the interval.}

\[
A^{\text{pep}}_{ij}
=
\begin{cases}
\dfrac{F_{ij}}{x_{t_0}(j)}, & x_{t_0}(j) > 0, \\[6pt]
\delta_{ij}, & x_{t_0}(j) = 0 ,
\end{cases}
\]

By construction, 
$A^{\text{pep}}$
 is column-stochastic on the same node set as 
$A^{\text{prod}}$, ensuring that discrepancies reflect estimation noise rather than normalization artifacts.

\subsection{Comparison metrics}

To quantify agreement between the two matrices, we evaluate matrix-level and column-level
discrepancies. Let
\[
D = A^{\text{pep}} - A^{\text{prod}} .
\]
We report a small set of standard, unweighted and population-weighted discrepancy measures to summarize agreement at both the global matrix level and the origin-column level.
\begin{itemize}
  \item the $\ell_1$ matrix norm $\|D\|_1 = \sum_{i,j} |D_{ij}|$,
  \item the root-mean-square error $\sqrt{\tfrac{1}{N^2}\sum_{i,j} D_{ij}^2}$,
  \item the mean column-wise $\ell_1$ deviation,
  \item and the mean column-wise Jensen--Shannon divergence.
\end{itemize}

Because some origin locations contribute substantially more trajectories than others, we also
compute population-weighted versions of these metrics. Using the normalized origin mass
\[
w_j = \frac{x_{t_0}(j)}{\sum_k x_{t_0}(k)},
\]
we define weighted column-wise deviations by averaging with respect to $w_j$. Weighted
$\ell_1$, RMSE, and Jensen--Shannon scores therefore emphasize agreement in high-flow regions
of the network while downweighting statistically sparse origins.

Formal definitions of all discrepancy metrics used in this section are provided
in Appendix~\ref{app:metrics}.

%\subsection{Interpretation}
%
%Agreement between $A^{\text{pep}}$ and $A^{\text{prod}}$ confirms that realized PEP trajectories
%faithfully instantiate the prescribed Markov dynamics at the aggregate level. Any discrepancy is
%attributable solely to finite-population sampling noise rather than to inconsistencies in the
%construction of the transition matrices.

\subsection{Verification results}

We now report a quantitative verification of the network-driven trajectory
generator. 

\subsubsection{Verification setup.}
The experiment uses the corridor overlay network on the H3 grid at
resolution~6, with single-feeder mode, potential-based edge directionality, and
the periodic fixed-point initialization \(\mathbf{p}_0=\mathbf{p}^\ast\). We
consider a representative verification window consisting of six consecutive
30-minute time-steps (06{:}00--09{:}00) on 2025--06--01. All runs use the same
network (99 nodes) and differ only in the number of realized PEP trajectories
\(K\).

\begin{table}[H]
\centering
\begin{tabular}{l l}
\hline
Parameter & Value \\
\hline
Region & Atlanta, Georgia \\
Graph mode & Corridors \\
H3 resolution & 6 \\
time-step size & 30 minutes \\
Verification span start & 2025--06--01 06{:}00 \\
Verification span end & 2025--06--01 09{:}00 \\
Number of time-steps & 6 \\
Number of nodes & 99 \\
\hline
\end{tabular}
\caption{Common configuration parameters for the flow verification experiments.
Only the number of realized PEP trajectories \(K\) varies across runs.}
\label{tab:verification-config}
\end{table}

\subsubsection{Quantitative results.}
Table~\ref{tab:verification-metrics-vsK} reports agreement between the
empirical end-to-end transition matrix \(A^{\text{pep}}\) estimated from
realized PEP trajectories and the matrix \(A^{\text{prod}}\) obtained by
composing the prescribed single time-step Markov matrices \(\{M_t\}\)
over the same verification window.

Unweighted metrics characterize global matrix-level discrepancies, while
population-weighted metrics emphasize agreement for origins contributing the
largest flow volumes. Because PEP trajectories are generated by independent
categorical draws from the columns of \(M_t\), discrepancies between
\(A^{\text{pep}}\) and \(A^{\text{prod}}\) arise solely from finite-sample
noise.

\begin{table}[H]
\centering
\begin{tabular}{l c c}
\hline
Metric & Run 1 (\(K=120{,}000\)) & Run 2 (\(K=480{,}000\)) \\
\hline
$\ell_1$ matrix norm & 12.4853 & 6.4588 \\
RMSE & $3.7399 \times 10^{-3}$ & $1.9336 \times 10^{-3}$ \\
Mean column $\ell_1$ deviation & 0.12611 & 0.06524 \\
Mean column Jensen--Shannon divergence &
$8.7402 \times 10^{-3}$ & $2.5204 \times 10^{-3}$ \\
Weighted $\ell_1$ deviation & 0.09497 & 0.04697 \\
Weighted RMSE & $2.7420 \times 10^{-3}$ & $1.3656 \times 10^{-3}$ \\
Weighted Jensen--Shannon divergence &
$5.1578 \times 10^{-3}$ & $1.4572 \times 10^{-3}$ \\
\hline
\end{tabular}
\caption{Agreement between the empirical end-to-end transition matrix
\(A^{\text{pep}}\) estimated from realized PEP trajectories and the
multi-step transition matrix \(A^{\text{prod}}\) obtained by composing the
prescribed single time-step Markov matrices \(\{M_t\}\), for two different numbers of realized PEP
trajectories \(K\).
All discrepancies arise solely from finite-population sampling noise.}
\label{tab:verification-metrics-vsK}
\end{table}
Across all metrics, discrepancies decrease as the number of realized PEP
trajectories \(K\) increases, consistent with finite-sample estimation of
transition probabilities. This behavior is expected for independent categorical
sampling from the Markov matrices, for which estimation error scales on the
order of  \(K^{-1/2}\).  The observed scaling is consistent with standard Monte Carlo estimation of categorical transition probabilities and confirms that discrepancies arise solely from finite-population sampling.

\subsubsection{Code availability.}
All network construction, trajectory generation, and verification procedures
described in this paper are implemented in a codebase available at
\url{https://github.com/asif-shakeel/Mobility-Trajectories.git}. The repository includes scripts to reproduce all figures and
verification metrics reported here.

\section{Summary and outlook}
Taken together, these results demonstrate that network-driven,
time-dependent Markov dynamics can be realized faithfully as synthetic trajectories of interchangeable movers without invoking agent memory or behavioral
assumptions. The framework provides a principled bridge between aggregated
OD flow descriptions and trajectory-level realizations, enabling the study
of time-elapsed mobility measures using only network structure and
externally specified schedules. Because all outputs are generated directly
from network-driven matrices, the model is particularly well suited to
privacy-preserving settings where individual-level data are unavailable.

The modular structure of the framework admits extension to settings in which  OD data are available for a specific city or region. Parameters governing network structure, temporal bias schedules, and distance decay may be adjusted or augmented to reflect observed aggregate flows, while preserving the separation between collective dynamics and individual-level behavior. Exploring such data-informed extensions is a natural direction for future work.

\appendix
\section{Perron--Frobenius theorem for the daily matrix}
\label{app:pf}

We briefly state the form of the Perron--Frobenius theorem relevant to the daily evolution matrix
used in this work.

Let \(Q \in \mathbb{R}^{N \times N}\) be a nonnegative, column-stochastic matrix, i.e.,
\(Q_{ij} \ge 0\) for all \(i,j\) and \(\sum_i Q_{ij} = 1\) for all \(j\).
Assume further that \(Q\) is irreducible, meaning that for any pair of indices \(i,j\), there exists
an integer \(k \ge 1\) such that \((Q^k)_{ij} > 0\).

Then the following statements hold:
\begin{itemize}
  \item The spectral radius of \(Q\) is equal to \(1\).
  \item There exists a unique (up to scalar normalization) strictly positive vector
        \(\mathbf{p}^\ast \in \mathbb{R}^N\) such that
        \[
        Q\,\mathbf{p}^\ast = \mathbf{p}^\ast .
        \]
  \item The vector \(\mathbf{p}^\ast\) can be normalized to represent a probability distribution,
        i.e., \(\mathbf{p}^\ast_i > 0\) and \(\sum_i \mathbf{p}^\ast_i = 1\).
\end{itemize}

In the present setting, the matrix \(Q = M_T \cdots M_1\) represents one full daily cycle of the
time-dependent Markov dynamics.
Irreducibility follows from the strong connectivity of the overlay mobility network together with the
assignment of strictly positive transition probability to every overlay edge at every time step:
for any pair of nodes, there exists a directed path in the overlay graph, and each edge along this
path is realized with positive probability at each step of the daily cycle.
Consequently, every node can reach every other node with positive probability over one full cycle.

Moreover, because each single time-step matrix \(M_t\) assigns strictly positive probability to
self-transitions, the daily matrix \(Q\) has a positive diagonal and is therefore primitive.
As a consequence, the invariant distribution \(\mathbf{p}^\ast\) is unique and globally attracting.

The resulting invariant distribution \(\mathbf{p}^\ast\) is thus a periodic fixed point of the
daily matrix and provides a natural, non-transient choice for the initial population distribution
\(\mathbf{p}_0\).

\section{Discrepancy metrics for matrix comparison}
\label{app:metrics}

This appendix defines the matrix- and column-level discrepancy metrics used to
compare the composed multi-step transition matrix \(A^{\text{prod}}\),
obtained by composing the prescribed single time-step transition matrices
\(\{M_t\}\), with the empirical matrix \(A^{\text{pep}}\) estimated from
realized PEP trajectories.

Let
\[
D = A^{\text{pep}} - A^{\text{prod}} \in \mathbb{R}^{N \times N}.
\]
Both matrices are column-stochastic matrices on the same node set.

\subsection*{Unweighted metrics}

\paragraph{Matrix $\ell_1$ norm:}
The matrix $\ell_1$ discrepancy is defined as
\[
\|D\|_1 = \sum_{i=1}^N \sum_{j=1}^N |D_{ij}|.
\]
This measures the total absolute deviation between the two matrices.

\paragraph{Root-mean-square error (RMSE):}
The RMSE is defined by
\[
\mathrm{RMSE}(D) =
\sqrt{\frac{1}{N^2} \sum_{i=1}^N \sum_{j=1}^N D_{ij}^2 }.
\]
This metric emphasizes large pointwise deviations while remaining scale-sensitive.

\paragraph{Mean column-wise $\ell_1$ deviation:}
For each origin column \(j\), define the column discrepancy
\[
\|D_{\cdot j}\|_1 = \sum_{i=1}^N |D_{ij}|.
\]
The reported metric is the average over columns,
\[
\frac{1}{N} \sum_{j=1}^N \|D_{\cdot j}\|_1 .
\]

\paragraph{Mean column-wise Jensen--Shannon divergence:}
For each column \(j\), let \(p_{\cdot j}\) and \(q_{\cdot j}\) denote the
\(j\)-th columns of \(A^{\text{pep}}\) and \(A^{\text{prod}}\), respectively,
interpreted as probability distributions.
The Jensen--Shannon divergence for column \(j\) is
\[
\mathrm{JS}(p_{\cdot j}, q_{\cdot j})
=
\frac{1}{2}
\mathrm{KL}\!\left(p_{\cdot j} \middle\| m_{\cdot j}\right)
+
\frac{1}{2}
\mathrm{KL}\!\left(q_{\cdot j} \middle\| m_{\cdot j}\right),
\]
where
\[
m_{\cdot j} = \tfrac{1}{2}(p_{\cdot j} + q_{\cdot j})
\]
and \(\mathrm{KL}(\cdot\|\cdot)\) denotes the Kullback--Leibler divergence.
For two discrete probability distributions \(p, q \in \mathbb{R}^N\) with
\(p_i \ge 0\), \(q_i \ge 0\), and \(\sum_i p_i = \sum_i q_i = 1\),
with the convention that terms with \(p_i = 0\) contribute zero.
The reported value is the mean over all columns.

\subsection*{Population-weighted metrics}

Because different origin nodes contribute different numbers of realized
trajectories, we also compute population-weighted versions of the column-wise
metrics. Let
\[
w_j = \frac{x_{t_0}(j)}{\sum_{k=1}^N x_{t_0}(k)},
\]
where \(x_{t_0}(j)\) is the number of PEP trajectories originating at node \(j\)
at the start of the verification window.

\paragraph{Weighted $\ell_1$ deviation:}
\[
\sum_{j=1}^N w_j \, \|D_{\cdot j}\|_1 .
\]

\paragraph{Weighted RMSE:}
\[
\sqrt{
\sum_{j=1}^N w_j
\left(
\frac{1}{N} \sum_{i=1}^N D_{ij}^2
\right)
}.
\]

\paragraph{Weighted Jensen--Shannon divergence:}
\[
\sum_{j=1}^N w_j \,
\mathrm{JS}(p_{\cdot j}, q_{\cdot j}).
\]

These weighted metrics emphasize agreement for origins that contribute the
largest fraction of realized flow and suppress noise arising from statistically
sparse columns.


\begin{bibdiv}
\begin{biblist}

\bib{bhg:slht}{article}{
author={Brockmann, D.},
author={Hufnagel, L.},
author={Geisel, T.},
title={The scaling laws of human travel},
journal={Nature},
volume={439},
number={7075},
pages={462--465},
year={2006},
note={\href{https://doi.org/10.1038/nature04292}{doi:10.1038/nature04292}}
}

\bib{ghb:uihmp}{article}{
author={Gonz{\'a}lez, M. C.},
author={Hidalgo, C. A.},
author={Barab{\'a}si, A.-L.},
title={Understanding individual human mobility patterns},
journal={Nature},
volume={453},
pages={779--782},
year={2008},
note={\href{https://doi.org/10.1038/nature06958}{doi:10.1038/nature06958}}
}

\bib{song2010modelling}{article}{
author={Song, C.},
author={Koren, T.},
author={Wang, P.},
author={Barab{\'a}si, A.-L.},
title={Modelling the scaling properties of human mobility},
journal={Nature Physics},
volume={6},
pages={818--823},
year={2010},
note={\href{https://doi.org/10.1038/nphys1760}{doi:10.1038/nphys1760}}
}

\bib{song2010}{article}{
author={Song, C.},
author={Qu, Z.},
author={Blumm, N.},
author={Barab{\'a}si, A.-L.},
title={Limits of predictability in human mobility},
journal={Science},
volume={327},
number={5968},
pages={1018--1021},
year={2010},
note={\href{https://doi.org/10.1126/science.1177170}{doi:10.1126/science.1177170}}
}

\bib{bbgj:hmma}{article}{
author={Barbosa, H.},
author={Barthelemy, M.},
author={Ghoshal, G.},
author={James, C. R.},
author={Lenormand, M.},
author={Louail, T.},
author={Menezes, R.},
author={Ramasco, J. J.},
author={Simini, F.},
author={Tomasini, M.},
title={Human mobility: Models and applications},
journal={Physics Reports},
volume={734},
pages={1--74},
year={2018},
note={\href{https://doi.org/10.1016/j.physrep.2018.01.001}{doi:10.1016/j.physrep.2018.01.001}}
}

\bib{simini2012}{article}{
author={Simini, F.},
author={Gonz{\'a}lez, M. C.},
author={Maritan, A.},
author={Barab{\'a}si, A.-L.},
title={A universal model for mobility and migration patterns},
journal={Nature},
volume={484},
pages={96--100},
year={2012},
note={\href{https://doi.org/10.1038/nature10856}{doi:10.1038/nature10856}}
}


\bib{foster2025pseudomarkov}{misc}{
author={Foster, A.},
author={Meyer, D. A.},
author={Shakeel, A.},
title={A Pseudo Markov-Chain Model and Time-Elapsed Measures of Mobility from Collective Data},
year={2025},
eprint={2502.04162},
archivePrefix={arXiv},
primaryClass={stat.AP},
note={\href{https://arxiv.org/abs/2502.04162}{arXiv:2502.04162}}
}

\bib{lwr1955}{article}{
author={Lighthill, M. J.},
author={Whitham, G. B.},
title={On kinematic waves. II. A theory of traffic flow on long crowded roads},
journal={Proceedings of the Royal Society of London. Series A},
volume={229},
number={1178},
pages={317--345},
year={1955},
note={\href{https://doi.org/10.1098/rspa.1955.0089}{doi:10.1098/rspa.1955.0089}}
}

\bib{richards1956}{article}{
author={Richards, P. I.},
title={Shock waves on the highway},
journal={Operations Research},
volume={4},
number={1},
pages={42--51},
year={1956},
note={\href{https://doi.org/10.1287/opre.4.1.42}{doi:10.1287/opre.4.1.42}}
}

\bib{daganzo1995}{article}{
author={Daganzo, C. F.},
title={The cell transmission model, part II: Network traffic},
journal={Transportation Research Part B: Methodological},
volume={29},
number={2},
pages={79--93},
year={1995},
note={\href{https://doi.org/10.1016/0191-2615(94)00022-R}{doi:10.1016/0191-2615(94)00022-R}}
}

\bib{bettencourt2013}{article}{
author={Bettencourt, L. M. A.},
title={The origins of scaling in cities},
journal={Science},
volume={340},
number={6139},
pages={1438--1441},
year={2013},
note={\href{https://doi.org/10.1126/science.1235823}{doi:10.1126/science.1235823}}
}

\bib{adnan2016simmobility}{inproceedings}{
author={Adnan, M.},
author={Pereira, F. C.},
author={Azevedo, C. M. L.},
author={Basak, K.},
author={Lovric, M.},
author={Raveau, S.},
author={Zhu, Y.},
author={Ferreira, J.},
author={Ben-Akiva, M.},
title={{SimMobility}: A multi-scale integrated agent-based simulation platform},
booktitle={Transportation Research Board 95th Annual Meeting Compendium of Papers},
year={2016},
note={TRB Paper No.~16-2691 (no DOI)}
}

\bib{krajzewicz2012sumo}{article}{
author={Krajzewicz, D.},
author={Erdmann, J.},
author={Behrisch, M.},
author={Bieker, L.},
title={Recent development and applications of SUMO -- simulation of urban mobility},
journal={International Journal on Advances in Systems and Measurements},
volume={5},
number={3--4},
pages={128--138},
year={2012}
}

\bib{horni2016matsim}{book}{
author={Horni, A.},
author={Nagel, K.},
author={Axhausen, K. W.},
title={The Multi-Agent Transport Simulation {MATSim}},
publisher={Ubiquity Press},
year={2016},
note={\href{https://doi.org/10.5334/baw}{doi:10.5334/baw}}
}

\bib{dgbcd:opcumpd}{article}{
author={de Montjoye, Y.-A.},
author={Gambs, S.},
author={Blondel, V. D.},
author={Canright, G.},
author={de Cordes, N.},
author={Deletaille, S.},
author={Eng{\o}-Monsen, K.},
author={Garcia-Herranz, M.},
author={Kendall, J.},
author={Kerry, C.},
author={Krings, G.},
author={Letouz{\'e}, E.},
author={Luengo-Oroz, M.},
author={Oliver, N.},
author={Rocher, L.},
author={Rutherford, A.},
author={Smoreda, Z.},
author={Steele, J.},
author={Wetter, E.},
author={Pentland, A.},
author={Bengtsson, L.},
title={On the privacy-conscientious use of mobile phone data},
journal={Scientific Data},
volume={5},
year={2018},
note={\href{https://doi.org/10.1038/sdata.2018.286}{doi:10.1038/sdata.2018.286}}
}

\bib{xtlzfj:trfa}{inproceedings}{
author={Xu, F.},
author={Tu, Z.},
author={Li, Y.},
author={Zhang, P.},
author={Fu, X.},
author={Jin, D.},
title={Trajectory Recovery From Ash: User Privacy Is NOT Preserved in Aggregated Mobility Data},
booktitle={Proceedings of the 26th International World Wide Web Conference},
pages={1241--1250},
year={2017},
note={\href{https://doi.org/10.1145/3038912.3052620}{doi:10.1145/3038912.3052620}}
}

\bib{tizzoni2014od}{article}{
author={Tizzoni, M.},
author={Bajardi, P.},
author={Poletto, C.},
author={Ramasco, J. J.},
author={Balcan, D.},
author={Gon{\c{c}}alves, B.},
author={Perra, N.},
author={Colizza, V.},
title={Real-time numerical forecast of global epidemic spreading: Case study of 2009 A/H1N1pdm},
journal={BMC Medicine},
volume={10},
pages={165},
year={2012},
note={\href{https://doi.org/10.1186/1741-7015-10-165}{doi:10.1186/1741-7015-10-165}}
}

\bib{cuebiqod}{misc}{
author={{Cuebiq}},
title={Cuebiq Mobility Data: Privacy-First Aggregated Origin--Destination Flows},
year={2020},
note={\href{https://www.cuebiq.com/}{cuebiq.com}}
}

\bib{safegraphod}{misc}{
author={{SafeGraph}},
title={SafeGraph Patterns and Origin--Destination Data},
year={2021},
note={\href{https://www.safegraph.com/}{safegraph.com}}
}

\bib{uberh3}{misc}{
author={{Uber Technologies, Inc.}},
title={H3: A Hexagonal Hierarchical Geospatial Indexing System},
year={2018},
note={\href{https://h3geo.org/}{https://h3geo.org/}}
}

\bib{dijkstra1959}{article}{
author={Dijkstra, E. W.},
title={A note on two problems in connexion with graphs},
journal={Numerische Mathematik},
volume={1},
number={1},
pages={269--271},
year={1959},
note={\href{https://doi.org/10.1007/BF01386390}{doi:10.1007/BF01386390}}
}

\bib{kwon2023multiple}{article}{
author={Kwon, Oh-Hyun},
author={Hong, Inho},
author={Jung, Woo-Sung},
author={Jo, Hang-Hyun},
title={Multiple gravity laws for human mobility within cities},
journal={EPJ Data Science},
volume={12},
pages={57},
year={2023},
note={\href{https://doi.org/10.1140/epjds/s13688-023-00438-x}{doi:10.1140/epjds/s13688-023-00438-x}}
}


\end{biblist}
\end{bibdiv}
\end{document}